# Realization of multiple charge density waves in NbTe$_2$ at the monolayer limit

Yusong Bai[1], Zemin Pan[1], Jinghao Deng[1], Xiaoyu Lin[1], Tao Jian[1], Chao Zhu[1], Da Huo[1], Zhengbo Cheng[1], Ping Cui[3], Zhenyu Zhang[3], Qiang Zou[2*], Chendong Zhang[1, 4*]

[1]*School of Physics and Technology, Wuhan University, Wuhan 430072, China*
[2]*Department of Physics and Astronomy, West Virginia University, WV 26506, USA*
[3]*International Center for Quantum Design of Functional Materials (ICQD), Hefei National Laboratory for Physical Sciences at the Microscale, University of Science and Technology of China, Hefei 230026, China*
[4]*Wuhan Institute of Quantum Technology, Wuhan 430206, China*

*Correspondence and requests for materials should be addressed to:
cdzhang@whu.edu.cn (C.D.Z.), qzou.iphy@gmail.com (Q.Z.)*

**Abstract: Layered transition-metal dichalcogenides (TMDCs) down to the monolayer (ML) limit provide a fertile platform for exploring charge-density waves (CDWs). Though bulk NbTe$_2$ is known to harbor a single axis 3 × 1 CDW coexisting with non-trivial quantum properties, the scenario in the ML limit is still experimentally unknown. In this study, we unveil the richness of the CDW phases in ML NbTe$_2$, where not only the theoretically predicted 4 × 4 and 4 × 1 phases, but also two unexpected $\sqrt{28} \times \sqrt{28}$ and $\sqrt{19} \times \sqrt{19}$ phases, can be realized. For such a complex CDW system, we establish an exhaustive growth phase diagram *via* systematic efforts in the material synthesis and scanning tunneling microscope characterization. Moreover, we report that the energetically stable phase is the larger scale order ($\sqrt{19} \times \sqrt{19}$), which is surprisingly in contradiction to the prior prediction (4 × 4). These findings are confirmed using two different kinetic pathways, *i.e.*, direct growth at proper growth temperatures ($T$), and low-$T$ growth followed by high-$T$ annealing. Our results provide a comprehensive diagram of the "zoo" of CDW orders in ML 1$T$-NbTe$_2$ for the first time and offer a new material platform for studying novel quantum phases in the 2D limit.**

Charge density waves (CDWs), a collective electronic phenomenon with atomic-scale periodic modulation in terms of both lattice and charge degrees of freedom,[1] have been discovered in many layered transition-metal dichalcogenides (TMDCs).[2,3] On reducing dimensionality, the electron–electron and electron–phonon interactions can be markedly enhanced due to the quenched screening,[2-4] particularly in the monolayer (ML) limit, where interactions in the vertical direction were further removed. As a consequence, exotic quantum phenomena associated with the new CDW orders emerge in such a ML system. For instance, ML VSe$_2$ shows a new symmetry-broken $\sqrt{7} \times \sqrt{3}$ charge order with an enhanced CDW transition temperature, in contrast to the 4 × 4 CDW in bulk VSe$_2$.[5,6]

The group V dichalcogenides MX$_2$ (M = V, Nb, Ta; X = S, Se, Te) are one of the best-known material families to harbor assorted 2D CDW orders.[7-12] Unlike other relatives in this family, niobium ditelluride (NbTe$_2$) has been less explored in terms of both experiment and theory. Previous experiments with bulk NbTe$_2$ have shown the coexistence of a 3 × 1 CDW order and superconductivity,[13-16] as well as anomalous magnetoresistance behaviors.[16,17] Owing to the shorter Te-Te spacing between layers, the interlayer interaction in Te-based TMDCs is believed to be stronger than that in Se/S compounds.[18] Therefore, the telluride layered materials could show dramatic thickness-dependent structural, electronic, magnetic, and topological phase transitions.[19–22] Recently, pioneering theoretical effort has predicted that the ML NbTe$_2$ may exhibit 4 × 4, 4 × 1, 3 × 3, and 3 × 1 CDW orders. The 4 × 4 phase is the most stable among them, though the differences in their formation energies are fairly small.[23] However, despite successes in materials fabrication,[24,25] the charge density modulations in NbTe$_2$ at the ML limit remain experimentally unexplored.

Herein, we report the successful epitaxial growth of ML NbTe$_2$ on a bilayer graphene/SiC substrate. Using scanning tunneling microscopy/spectroscopy (STM/S), the triaxial 4 × 4 and uniaxial 4 × 1 CDW, which are absent in the bulk NbTe$_2$, are found in the ML limit agreeing with the calculations.[23] For both phases, a correlation gap with remarkable magnitude (or an asymmetric line shape in the vicinity of Fermi level) and the intensity reversal of local density-of-state (DOS) maps at opposite biases are observed, which strongly support their CDW origin. Moreover, we discover unexpected triaxial CDW orders with much larger periodicities (*i.e.*, $\sqrt{28} \times \sqrt{28}$ and $\sqrt{19} \times \sqrt{19}$). *Via* systematic efforts in the synthesis and the STM characterization, we establish a comprehensive growth diagram for such a complex CDW system. Confirmed by using two different kinetic pathways, we conclude that the theoretically unforeseen incommensurate CDW orders are generally more thermally favored. The pure phase of $\sqrt{19} \times \sqrt{19}$ CDW is attained at substantially high growth (or annealing) temperature (> 400 °C), manifesting itself as the most energetically stable CDW in ML NbTe$_2$. The successful controlled growth of multiple types of CDW orders in ML 1$T$-NbTe$_2$ provides a prospective material system for exploring novel quantum phases in the 2D limit.

**Results and discussion**

ML NbTe$_2$ is in a high-symmetry octahedral (1$T$) structure at room temperature,[24,25] where the hexagonally arranged Nb atoms are sandwiched by two layers of Te atoms in an octahedral coordination, as sketched in Figure 1a (top view in the upper panel and side view in the lower panel). Using molecular beam epitaxy (MBE), we successfully grew ML NbTe$_2$ on a bilayer graphene (BLG)-terminated 6H-SiC(0001) substrate. Figure 1b shows a typical large-scale STM image of sub-ML NbTe$_2$. The apparent height of the ML NbTe$_2$ film was 8.9 Å, as shown in the inset of Figure 1b. The room-temperature atomically resolved STM image (Figure 1c) clearly shows a well-ordered hexagonal lattice with a lattice constant of $a = 3.64 \pm 0.04$ Å. *In situ* reflection high-energy electron diffraction (RHEED) patterns were measured to monitor the growth process of NbTe$_2$ on BLG. After 1 hour of growth (0.9 ML coverage), two sets of strikes appeared (Figure 1d), which suggests that NbTe$_2$ has its own lattice structure rather than taking on the lattice structure of BLG. Using BLG as a reference, the lattice constant of ML NbTe$_2$ was estimated to be $a = 3.68$ Å, which is consistent with both our STM results (Figure 1c) and the reported value of 3.66 Å for NbTe$_2$ nanoplates.[26] X-ray photoemission spectroscopy (XPS) measurements were performed to characterize the formation of NbTe$_2$. As shown in Figure S1, the Nb 3d$^{5/2}$ (201.9 eV) and Nb 3d$^{3/2}$ (204.7 eV) peaks from Nb$^{4+}$, as well as the Te 3d$^{5/2}$ (573.2 eV) and Te 3d$^{3/2}$ (581.4 eV) peaks from Te$^{2-}$, were detected, which is consistent with previous work.[27] The combined STM, RHEED, and XPS characterization results confirm the successful epitaxial growth of a high-quality ML NbTe$_2$ film on a BLG-terminated 6H-SiC(0001) substrate.

We fabricated ML NbTe$_2$ with a series of growth temperatures ($T_{growth}$) and characterized those samples at 4.7 K in an MBE-STM integrated system. Figure 2a shows a 30 × 30 nm$^2$ STM image of 0.8 ML NbTe$_2$ with $T_{growth}$ = 250 °C that displays the coexistence of three superstructures: (1) a well-ordered hexagonal superstructure with a periodicity of 1.45 nm (marked by the solid cyan rhombus and labeled as a 4 ×

4 superstructure); (2) a 1D striped superstructure with a periodicity of 1.28 nm along the direction perpendicular to the stripes (labeled as a 4 × 1 superstructure); and (3) a disordered hexagonal superstructure that exhibits a local triaxial period but without long-range order (marked by the dashed blue rhombus and labeled as a disordered $\sqrt{28} \times \sqrt{28}$ superstructure). Increasing the growth temperature to 300 °C resulted in more 4 × 1 phase and ordered $\sqrt{28} \times \sqrt{28}$ phase with a periodicity of 1.91 nm, but the 4 × 4 phase disappeared (Figure 2b and Figure S2). Interestingly, the 4 × 1 phase was not found after the $T_{growth}$ was increased to 350 °C, while the $\sqrt{28} \times \sqrt{28}$ phase continued to increase and coexisted with a new hexagonal superstructure with a periodicity of 1.58 nm (marked by the solid red rhombus and labeled as a $\sqrt{19} \times \sqrt{19}$ superstructure in Figure 2c and Figure S2). Further increasing the substrate temperature (400-450 °C) yielded a pure $\sqrt{19} \times \sqrt{19}$ phase (Figure 2d). We summarized the evolution of the four superstructures in ML NbTe$_2$ as a function of growth temperature in Figure 2e. The surface morphology evolution is shown to be driven by $T_{growth}$, indicating that the 4 × 4, 4 × 1 and $\sqrt{28} \times \sqrt{28}$ phases are metastable, while the $\sqrt{19} \times \sqrt{19}$ phase is energy stable.

To examine the phase stability of those superstructures, we carried out a post-annealing process on the low-temperature grown ML NbTe$_2$. Figure 3a shows an STM image of ML NbTe$_2$ grown and annealed at 250 °C, which is in good agreement with the results in Figure 2a. Different from the direct growth case, the 4 × 1 superstructure was not observed when we fully annealed the sample to 300 °C (Figure 3b, more results are seen in Figure S3). This suggests the Te-rich conditions may be essential to prepare the 4 × 1 phase (by lowering the formation enthalpy). Further increasing the annealing temperature to 350 °C led to a higher coverage of the $\sqrt{28} \times \sqrt{28}$ superstructure and an appearance of the $\sqrt{19} \times \sqrt{19}$ superstructure (Figure 3c), same as Figure 2c. A pure $\sqrt{19} \times \sqrt{19}$ phase was generated after the film was annealed over 400 °C (Figure 3d), similar to the results of the direct growth cases with growth temperatures from 400 to 450 °C (Figure 2d). The evolution of those superstructures upon post-annealing confirms the instability of the 4 × 4, 4 × 1 and $\sqrt{28} \times \sqrt{28}$ phases, which was further verified by another post-annealing process on the sample grown at 300 °C (see Figure S3 for details).

In ref. 23, first-principles calculations suggested the appearance of 4 × 4, 4 × 1, 3 × 3, and 3 × 1 CDW orders. Our experiments realized the predicted 4 × 4 and 4 × 1 superstructures, but not the 3 × 3 and 3 × 1 orders. To validate their CDW origins, we performed further STM/S investigations on those two phases. Figure 4a shows an atomic-resolved STM image of the 4 × 1 superstructure. Remarkably, a $4a$ periodicity (~1.45 nm) along the direction of the atomic lattice can be distinguished (Figure 4b). This $4a$ modulation is confirmed by the two additional peaks circled in green in the fast-Fourier transform (FFT) image (Figure 4c). Moreover, we noticed that the distortion of the topmost Te atoms was significantly large (see Figure 4a). To give a more quantitative description of the lattice distortion, we labeled the four atoms in the $4a$ period as $Te_1$, $Te_2$, $Te_3$, and $Te_4$ (marked in Figure 4b) and summarized the distance between adjacent Te atoms in Table S1. It was found that the distance between $Te_1$-$Te_2$ and $Te_4$-$Te_1$ ($Te_2$-$Te_3$ and $Te_3$-$Te_4$) was longer (shorter) than the Te-Te distance of an ideal octahedral structure, and the variation in the Te-Te bond length was up to 20%, which is much larger than the 1-7% distortion in conventional ML TMDs within the CDW order.[3,28]

We further examined the electronic structures of the 4 × 1 superstructure by STS. The large-scale tunneling spectra show a suppression of DOS around the Fermi level and display spatial modulation (homogeneity) in the direction perpendicular (parallel) to the stripes (see Figure S4). The magnified averaged *dI/dV* spectroscopy is shown in Figure 4d, where a soft gap of ~20 meV at the Fermi level was observed. This gap size is comparable to that of other typical CDW TMDs at the ML limit.[9,11,29] Figure 4e and 4f show the *dI/dV* maps of the same area obtained at -30 mV and +30 mV. The stripe modulations with an apparent intensity reversal are distinguished (see Figure S5 for more *dI/dV* maps). Generally, a spatial phase flip in the conductance maps at opposite energies near the Fermi level is a hallmark of the classic Peierls CDW, for which the intensity of the charge accumulation region is enhanced under negative bias, whereas the charge depletion region shows enhanced intensity under positive bias. Hence, the contrast reversal between the filled and empty states in the 4 × 1 phase strongly suggests that it originates from a CDW order.[30–32] Thus, the 4$a$ periodical modulation and atomic lattice distortion, the gap structure around the Fermi level, and the spatial phase flip in the conductance maps collectively support a CDW nature of the 4 × 1 superstructure

Similar CDW behaviors are also observed for the 4 × 4 superstructure. Figure 5a shows an STM image with atomic resolution, which captures a 1 × 1 hexagonal unit (marked in white) and 4 × 4 reconstructed patterns (marked in cyan). The line profile along the black arrow in Figure 5a displays 4*a* modulation, as shown in Figure 5b. The FFT image is shown in Figure 5c, displaying alignment between the 1 × 1 lattice peaks (marked in white circles) and the 4 × 4 spots (marked in cyan circles). In Figure 5d, the tunneling spectrum also shows a suppression of DOS near the Fermi level and the magnified *dI/dV* curve exhibits a particle-hole asymmetric V-shape gap (inset of Figure 5d and more spectroscopic results are seen in Figure S6). Such a spectroscopic feature (*i.e.* lack of sharp edge) is also commonly seen in other CDW systems like the $CsV_3Sb_5$ and $VSe_2$, presumably due to the partially gapped Fermi surface.[33,34] Moreover, the *dI/dV* maps of the same area measured at opposite biases exhibit intensity reversal as well. As indicated by the dashed cyan lines superimposed on Figure 5e and 5f, regions with the maximum charge intensity at -150 mV (Figure 5f) turn to the lowest charge intensity at +150 mV (Figure 5e). These results thus support the CDW nature of the 4 × 4 superstructure.

Our experiments not only confirm the 4 × 1 and 4 × 4 predictions, but also reveal the much larger-scale superstructures with irrational periodicities (namely, the $\sqrt{28} \times \sqrt{28}$ and $\sqrt{19} \times \sqrt{19}$) that are more stable, including the most energetically stable phase ($\sqrt{19} \times \sqrt{19}$). Noticed that the $\sqrt{19} \times \sqrt{19}$ structure was recently observed in the ML 1*T*-TaTe$_2$, showing nearly identical STM morphology features with ours, and has been unambiguously confirmed as an enlarged David-star CDW.[13,35] The $\sqrt{28} \times \sqrt{28}$ structure was also found to be a CDW order according to our private communications with Zeng. C. G *et, al*. Given the much larger length scales in these two phases, more defective features will occur naturally within the periodic unit cells. Thus, more efforts are needed to fully optimize their atomic structures and electronic properties. It is a demanding task to be carried out in future, but beyond the present scope. Despite the discrepancy in experiments and calculations,[23] it is for certain that the ML-NbTe$_2$ is endowed with rich and diversified electron/phonon correlation modes, which are greatly different from the bulk case.[13,14] In bulk NbTe$_2$, the interlayer Te-Te interaction is known to be rather strong which can even result in additional intralayer chalcogen-to-metal transfer (1/3$e$ per unit) forming the trimerized Nb ions chains.[18,36] When the thickness is down to a ML limit, the absent interlayer interaction and enhanced electron correlation may suppress the movement of electrons, leading to the emergence of new charge orders.

## Conclusion

In conclusion, we reported the MBE growth and STM/S characterization of NbTe$_2$ thin films down to the ML limit. Multiple CDW orders are found, which are entirely different from the known bulk case and deviate from the recent theoretical predictions as well.[13,14,23] In particular, we first confirmed the 4 × 4 and 4 × 1 CDWs predictions by thorough spectroscopic measurements. Moreover, we revealed unexpected CDW orders with larger length scales, including the $\sqrt{28} \times \sqrt{28}$ and the $\sqrt{19} \times \sqrt{19}$ that is, surprisingly, the stable phase. Precise control of the complex of multiple CDW phases can be achieved owing to the establishment of a comprehensive growth phase diagram of such an uncharged monolayer TMDC material. Our findings provide a promising platform for exploring novel properties associated with emerging CDW orders, such as nontrivial topology and magnetism.

## Methods

**Growth of the ML NbTe₂ films.** Sample growth was carried out using a home-built MBE system with a base pressure of ~1.2 × 10⁻¹⁰ Torr. The 6H-SiC(0001) wafer was first degassed at 650 °C for several hours and then flashed to 1300 °C for 45 cycles to obtain the bilayer graphene-terminated surface.[37] High-purity Nb (99.5%) and Te (99.999%) were evaporated from an electron-beam evaporator and a standard Knudsen cell, respectively. The flux ratio between Nb and Te was ~ 1:20. The substrate was heated by means of direct current and calibrated with an infrared spectrometer. The growth process was monitored by *in situ* RHEED operated at 20 kV. For *ex situ* STM/S measurements, a vacuum vessel with a base pressure of ~ 5 × 10⁻¹⁰ Torr was used to protect the sample from oxidation during the transition.

**Scanning tunneling microscopy and spectroscopy.** The STM/S measurements were carried out using a commercial *Unisoku* 1300 LT-STM system at 4.8 K (base pressure: < 1 × 10⁻¹⁰ Torr). Electrochemically etched tungsten tips were cleaned *in situ* with electron-beam bombardment and calibrated on a clean Cu(111) surface before all measurements. The *dI/dV* spectra were obtained at a constant tip-sample distance by using a standard lock-in technique with a modulation voltage of 973 Hz.

**X-ray photoelectron spectroscopy.** XPS spectra were acquired using a Thermo Fisher ESCALAB 250Xi instrument employing a monochromatic Mg K Alpha (1.254 keV) X-ray source in an ultrahigh vacuum atmosphere. The binding energies were calibrated using the C(1s) carbon peak (284.8 eV).

## Acknowledgements

This work was supported by the National Key R&D Program of China (2018FYA0305800 and 2018YFA0703700), the National Natural Science Foundation of China (11974012 and 12134011), and the Strategic Priority Research Program of Chinese Academy of Sciences (XDB30000000). We also thank Dr. Yong Liu from Wuhan University for the XPS measurements.

**Electronic Supplementary Material:** Supplementary material (including XPS results, STM images of the $\sqrt{28} \times \sqrt{28}$ and $\sqrt{19} \times \sqrt{19}$ CDW orders, large-scale tunneling spectra of the 4 × 1 and 4 × 4 phases, *dI/dV* maps of the 4 × 1 phase) is available in the online version of this article at http://XXX.

**Figure 1**

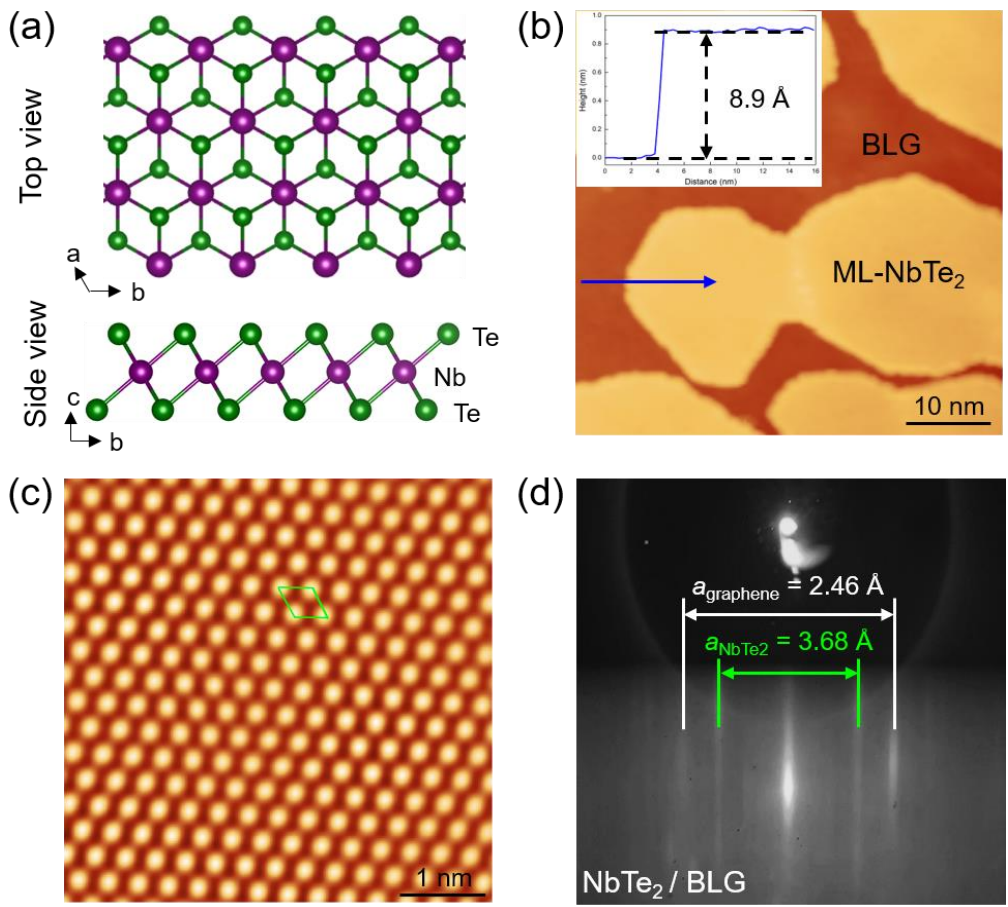

**Figure 1. Characterization of epitaxially grown ML NbTe$_2$.** (a) Crystal structure of monolayer 1*T*-NbTe$_2$. (b) Large-scale STM image of monolayer NbTe$_2$. ($V_{bias}$ = 3 V, $I_t$ = 10 pA). The inset shows the height profile along the blue arrow in (b). (c) Room-temperature atomically resolved STM image. ($V_{bias}$ = 10 mV, $I_t$ = 1 nA). It shows a hexagonal lattice with a periodicity of 3.64 ± 0.04 Å. (d) RHEED image of submonolayer NbTe$_2$ on a BLG substrate.

# Figure 2

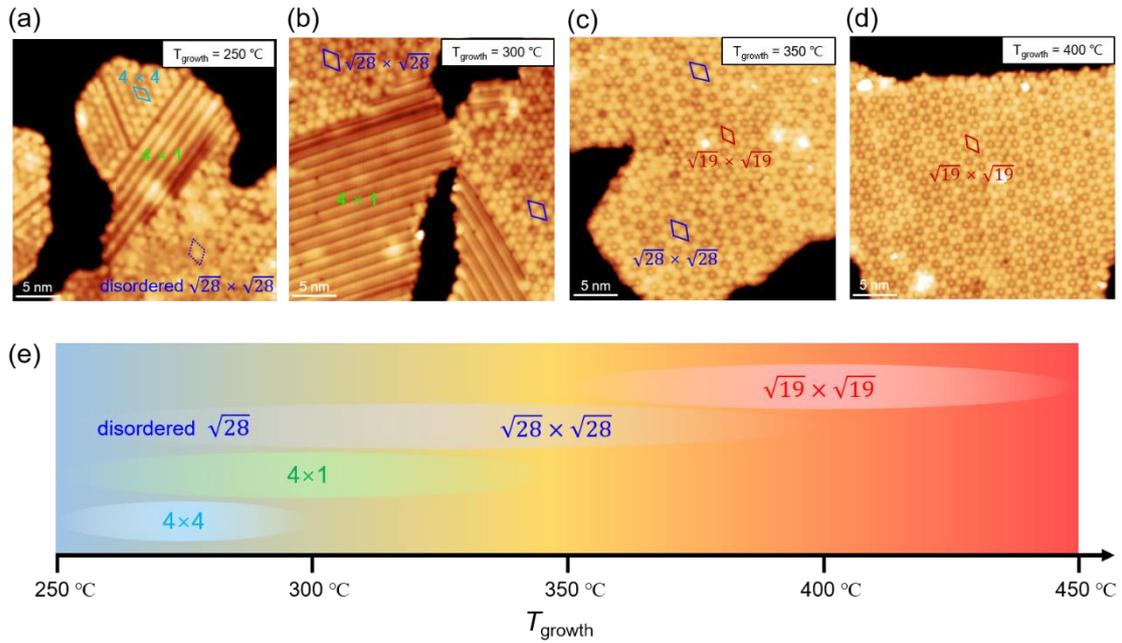

**Figure 2. Control of four superstructures in ML NbTe$_2$ by growth temperature.** (a-d) STM topographies of sub-ML NbTe$_2$ at growth temperatures of (a) 250, (b) 300, (c) 350, and (d) 400 °C. Each superstructure is marked in the Figure. (e) Phase diagram as a function of growth temperature. Scanning parameters for all images: $V_{bias}$ = -1 V, $I_t$ = 10 pA. All images were obtained at 4.7 K.

**Figure 3**

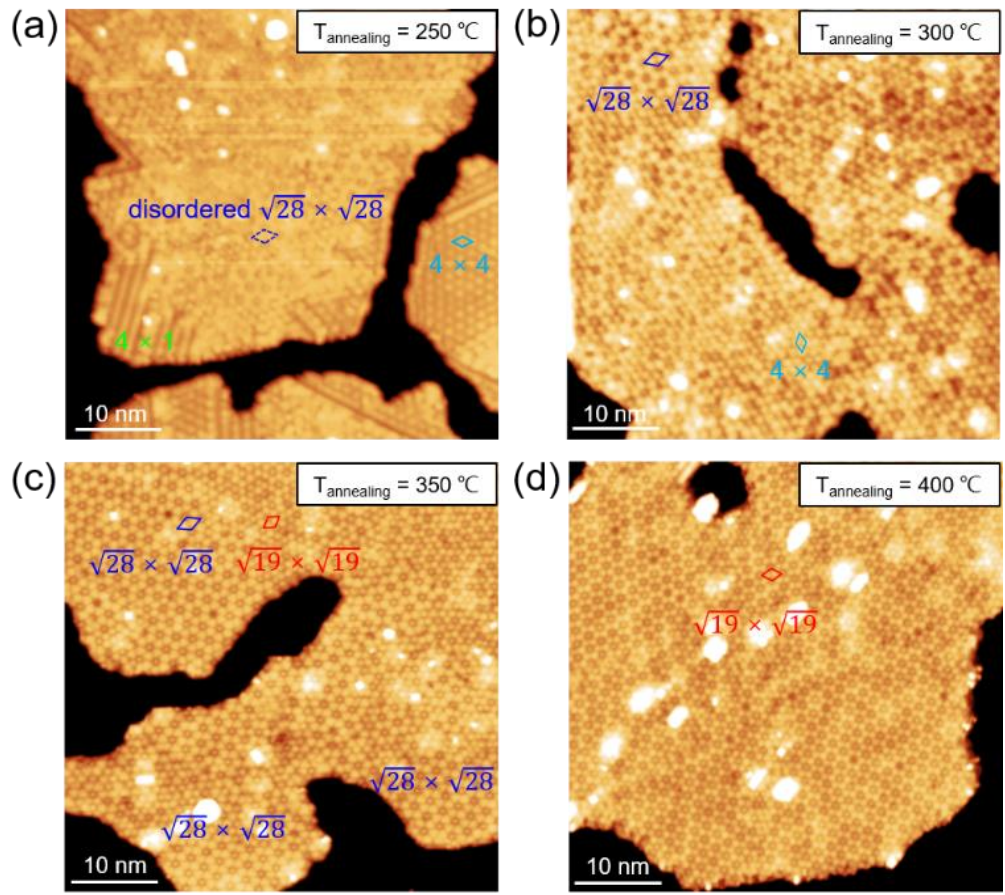

**Figure 3. Phase transition induced by post-annealing in ML NbTe$_2$.** (a-d) STM images of sub-ML NbTe$_2$ with *in situ* vacuum annealing at (a) 250, (b) 300, (c) 350, and (d) 400 °C for 1 hour. Each superstructure is marked in the Figure. Scanning parameters for all images: $V_{bias}$ = -1 V, $I_t$ = 10 pA.

# Figure 4

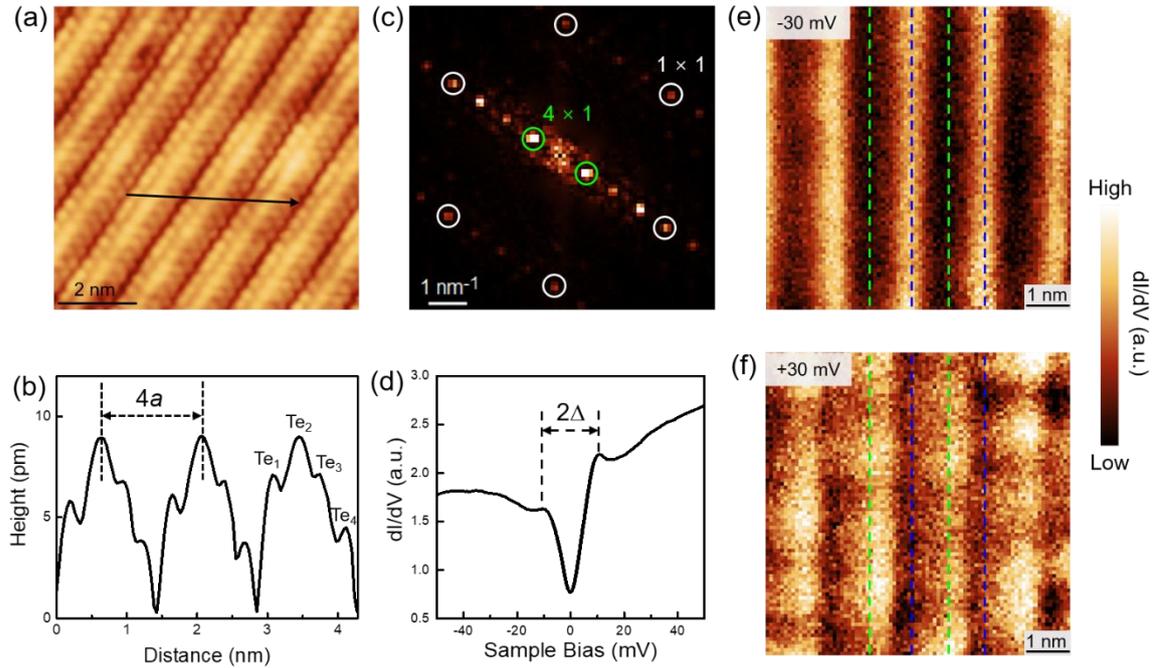

**Figure 4. STM/S characterization of the 4 × 1 CDW phase.** (a) Atomically resolved STM image of the 4 × 1 CDW phase. ($V_{bias}$ = 100 mV, $I_t$ = 100 pA). (b) Line profile across the NbTe$_2$ atomic lattice, shown by the black arrow in (a). (c) Fast-Fourier transform image of (a), where six 1 × 1 spots (white circles) and a pair of 4 × 1 CDW spots (green circles) can be clearly seen. (d) Typical tunneling spectrum of the 4 × 1 CDW phase. ($V_{bias}$ = 50 mV, $I_t$ = 200 pA). (e) Conductance maps obtained at E = -30 mV and (f) E = +30 mV over the same region show the stripe modulations with a clear intensity reversal. The green and blue dashed lines are visual guides. All the data were obtained at 4.7 K.

**Figure 5**

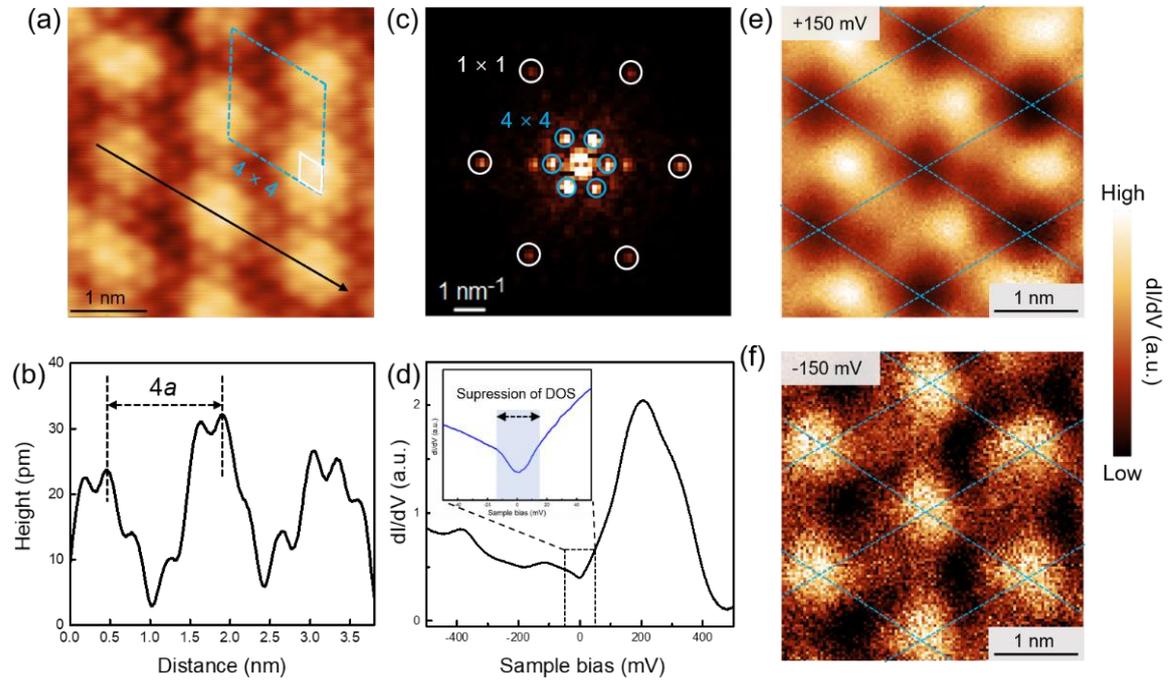

**Figure 5. STM/S characterization of the 4 × 4 CDW phase.** (a) Atomically resolved STM image of the 4 × 4 CDW phase. ($V_{bias}$ = -100 mV, $I_t$ = 5 nA). (b) Line profile across the black arrow in (a) shows the 4$a$ modulation. (c) Fast-Fourier transform image of (a). The white and cyan circles indicate the peaks associated with the Bragg points and 4$a$ × 4$a$ modulation, respectively. (d) Typical large-scale tunneling spectrum of the 4 × 4 CDW phase. ($V_{bias}$ = -500 mV, $I_t$ = 200 pA). Inset: magnified $dI/dV$ spectroscopy exhibiting a suppression of the DOS near the Fermi level. ($V_{bias}$ = 50 mV, $I_t$ = 200 pA). (e) Conductance maps of the same area obtained at E = +150 mV and (f) E = -150 mV. A clear charge modulation with contrast inversion can be clearly seen in these maps. The cyan dashed lines are visual guides.